\documentstyle[12pt]{article}
\makeatletter
\def\thesection{\arabic{section}.}
\@addtoreset{equation}{section}

\def\appendix{\par
\setcounter{section}{0}
\setcounter{subsection}{0}
\def\thesection{\Alph{section}.}}
\makeatother
\makeatletter
\def\abstract#1{\long\def\@abstract{#1}}%
\def\@abstract{}%
\let\@oldmaketitle=\@maketitle%
\def\@maketitle{%
\@oldmaketitle%
\begin{center}\large\bf Abstract\end{center}%
\begin{quotation}\@abstract\end{quotation}%
\vskip 1.5em}%
\makeatother
\newcommand{\delphi}{\Delta\varphi}
\newcommand{\delt}{\Delta t}
\newcommand{\tr}{{\rm tr}}
\begin{document}

\title{\sl More about Path Integral for Spin}
\author{
Kunio FUNAHASHI\thanks{e-mail: fnhs1scp@mbox.nc.kyushu-u.ac.jp,
fax: +81-92-633-4525},
Taro KASHIWA\thanks{e-mail: taro1scp@mbox.nc.kyushu-u.ac.jp}\\
Department of Physics, Kyushu University, Fukuoka 812-81, Japan\\\\
Shuji NIMA\thanks{e-mail: h79137g@kyu-cc.cc.kyushu-u.ac.jp,
fax: +81-93-692-3245}\\
Kyushu Women's Junior College, Kitakyushu 807, Japan\\\\
Seiji SAKODA\thanks{e-mail: a00501@cc.hc.keio.ac.jp}\\
Department of Physics, Hiyoshi, Keio University,\\
Yokohama 233, Japan}
\date{\today}
\abstract{Path integral for the $SU(2)$ spin system is reconsidered.
We show that the Nielsen-Rohrlich(NR) formula is equivalent
to the spin coherent state expression so that the phase space in the
NR formalism is not topologically nontrivial. We also perform the WKB
approximation in the NR formula and find that it gives the exact result.}
\maketitle\thispagestyle{empty}
\newpage

\section{Introduction}

It is known that there are at least two path integral expressions for the
$SU(2)$ spin system: one is the Nielsen-Rohrlich formula\cite{NR} and
the other is expressed by a spin coherent state\cite{KS,AP}.
The kernel in the Nielsen-Rohrlich (NR) formula, when Hamiltonian is
given by $H(t)=h(t)J_3$, with $h(t)$ being an external magnetic field,
is read as
\begin{eqnarray}
K\left(\varphi_F,t_F;\varphi_I,t_I\right)\!\!&=&\!\!\sum^\infty_{n=-\infty}
\lim_{N\to\infty}\int^\infty_{-\infty}\prod^{N-1}_{i=1}
{d\varphi_i\over2\pi}
\int^{J+1/2-\varepsilon}_{-J-1/2+\varepsilon}\prod^N_{j=1}{dp_j\over2\pi}
\label{nrf}\\
&&\left.\times\exp\left[ i\sum^N_{k=1}\left( p_k+J\right)\delphi_k
-\delt\sum^N_{k=1}h\!\left( k\right)p_j\right]
\right\vert^{\varphi_N=\varphi_F+2n\pi}_{\varphi_0=\varphi_I}\ ,\nonumber
\end{eqnarray}
where $\varepsilon$ is positively infinitesimal,
\begin{equation}
\delphi_j\equiv\varphi_j-\varphi_{j-1}\ ,\label{delphiteigi}
\end{equation}
and
\begin{equation}
\delt\equiv\left( t_F-t_I\right)/N\ .\label{deltteigi}
\end{equation}
In this expression, because of the existence of infinite sum
(and the integration range of $p_j$'s,
see the following for details), it is believed that the phase space is
nontrivial: ``punctured sphere''.

On the other hand the same system is expressed by means of a spin coherent
state:
the kernel is given as
\begin{eqnarray}
K\left(\xi_F,t_F;\xi_I,t_I\right)&=&\lim_{N\to\infty}\int\prod^{N-1}_{j=1}
{2J+1\over\pi}{d\xi_j^*d\xi_j\over\left(1+\vert\xi_j\vert^2\right)^2}
\exp\Bigg[ J\sum^N_{k=1}\Bigg\{2\log{1+\xi_k^*\xi_{k-1}\over
1+\vert\xi_k\vert^2}\\
&&-i\delt h\!\left( k\right){1-\xi_k^*\xi_{k-1}\over1+\xi_k^*\xi_{k-1}}\Bigg\}
\left.+\log\left(1+\vert\xi_N\vert^2\right)
-\log\left(1+\vert\xi_0\vert^2
\right)\Bigg]\right\vert^{\xi_N=\xi_F}_{\xi_0=\xi_I}\ .\nonumber
\end{eqnarray}
In this case there appears no infinite sum and the phase space is
$CP^1\simeq S^2$: ``sphere''.
Therefore these two formulae describing the same $SU(2)$ spin system look
very different.
We wish to know the origin of this difference, which is a main motivation
in this paper.

Also as for the WKB approximation there seems a discrepancy:
the former one gives us a equation of motion
\begin{equation}
\cases{\displaystyle\dot\varphi\left( t\right)=h\!\left( t\right)\ ,\cr
\displaystyle\dot p\left( t\right)=0\ ,\cr}\label{eqm}
\end{equation}
which apparently does not meet the boundary condition, $\varphi(T)=\varphi_F$
and $\varphi(0)=\varphi_I$, so that {\em there seems no classical solution},
while in the latter case the WKB approximation
yields the exact result\cite{FKSF}.
Therefore it needs to study this issue, too.

The plan of this paper is as follows: in section 2 these two formulae are
explicitly constructed by use of coherent states.
The next section 3 explains the origin of the winding number in $S^1$ case,
which clarifies the nature of the ``winding number'' in the NR formula.
Then in section 4 we show an explicit connection of both formulae so that
the WKB-exactness should hold also in the NR formula, which is the subject
of section 5.
The final section 6 is devoted to discussion and some mathematical relations
needed for the proof of the WKB-exactness are shown in the appendix.

\section{Path Integral Formulae for Spin}

Hamiltonian is supposed as
\begin{equation}
H(t)=h_3\!(t)J_3+h_-\!(t)J_++h_+\!(t)J_-\ ,\label{hamiltonian}
\end{equation}
where $J$'s are the generators of $SU(2)$ satisfying
\begin{equation}
\left[ J_+,J_-\right]=2J_3\ ,\ \left[ J_3,J_\pm\right]=\pm J_\pm\ ,\
\left( J_\pm\equiv J_1\pm iJ_2\right)\ ,
\end{equation}
and $h$'s are time dependent external fields with
\begin{equation}
h_\pm\!(t)\equiv
{1\over2}\left(h_1\!\left( t\right)\pm ih_2\!\left( t\right)\right)\ .
\end{equation}
We adopt the spin $J$ representation:
\begin{eqnarray}
&&J^2\vert m\rangle=J\left( J+1\right)\vert m\rangle\ ;
\ \left( J^2\equiv{J_1}^2+{J_2}^2+{J_3}^2\right)\ ,\nonumber\\
&&J_3\vert m\rangle=\left( m-J\right)\vert m\rangle\ ;
\ \left(0\le m\le2J\right)\ ,\nonumber\\
&&J_+\vert m\rangle=\sqrt{\left(2J-m\right)\left( m+1\right)}
\vert m+1\rangle\ ,\nonumber\\
&&J_-\vert m\rangle=\sqrt{m\left(2J-m+1\right)}\vert m-1\rangle\ .
\label{hyouji}
\end{eqnarray}

First construct the NR formula\cite{TK}:
to this end introduce the periodic coherent state,
\begin{equation}
\vert\varphi\rangle\equiv{1\over\sqrt{2\pi}}\sum^{2J}_{m=0}
e^{-im\varphi}\vert m\rangle\ ,\label{pcs}
\end{equation}
whose inner product is calculated to be
\begin{equation}
\langle\varphi\vert\varphi^\prime\rangle={1\over2\pi}\sum^{2J}_{m=0}
e^{im\delphi}\ ;\ \delphi\equiv\varphi-\varphi^\prime\ .\label{naiseki}
\end{equation}
For a later convenience, we introduce the following formula:
\begin{flushleft}{{\bf formula:}\ \  for $m_0,m_1\in{\bf Z}$}\end{flushleft}
\begin{equation}
\sum^{m_1}_{m=m_0}e^{im\varphi}f\!\left( m\right)=\sum^\infty_{n=-\infty}
\int^{m_1+\varepsilon_0}_{m_0-\varepsilon_0}dp\
e^{ip\left(\varphi+2n\pi\right)}
f\!\left( p\right)\ ;\
\left(0<\varepsilon_0<1\right)\ .\label{kankeishiki}
\end{equation}
Prove this: inserting a trivial integral to the left hand side to find
\begin{eqnarray}
\sum^{m_1}_{m=m_0}e^{im\varphi}f\!\left( m\right)&=&
\sum^\infty_{m=-\infty}\int^{m_1+\varepsilon_0}_{m_0-\varepsilon_0}dp\
\delta\!\left( p-m\right) e^{im\varphi}f\!\left( m\right)\nonumber\\
&=&\sum^\infty_{m=-\infty}\int^{m_1+\varepsilon_0}_{m_0-\varepsilon_0}dp\
\delta\!\left( p-m\right) e^{ip\varphi}f\!\left( p\right)\ ,
\end{eqnarray}
where it should be noted that $\varepsilon_0$ is needed
to avoid the $\delta$-function singularity at the upper as well as
the lower limit of the integration.
Applying the relation
\begin{equation}
\sum^\infty_{m=-\infty}\delta\!\left( p-m\right)=\sum^\infty_{n=-\infty}
e^{i2n\pi p}\ ,
\end{equation}
we arrive at the right hand side,
which completes the proof of (\ref{kankeishiki}).

Making use of the formula (\ref{kankeishiki}), that is, setting
$m_1=2J$ and $m_0=0$, we can rewrite (\ref{naiseki}) to
\begin{eqnarray}
\langle\varphi\vert\varphi^\prime\rangle&=&
\sum^\infty_{n=-\infty}\int^{2J+\varepsilon_0}_{-\varepsilon_0}{dp\over2\pi}
e^{ip\left(\delphi+2n\pi\right)}\nonumber\\
&\stackrel{p\mapsto p-J}{=}&\sum^\infty_{n=-\infty}
\int^{\lambda-\varepsilon}_{-\lambda+\varepsilon}{dp\over2\pi}
e^{i\left( p+J\right)\left(\delphi+2n\pi\right)}\ ,\label{phiphidash}
\end{eqnarray}
where $0<\varepsilon_0<1$ so that $-1/2<\varepsilon<1/2$ and
\begin{equation}
\lambda\equiv J+{1\over2}\ .\label{lteigi}
\end{equation}

The resolution of unity holds:
\begin{equation}
\int^{2\pi}_0d\varphi\vert\varphi\rangle\langle\varphi\vert={\bf 1}_J
\equiv\sum^{2J}_{m=0}\vert m\rangle\langle m\vert\ .
\label{phinoru}
\end{equation}
The matrix elements of generators are calculated in a similar manner such that
\begin{eqnarray}
\langle\varphi\vert J_3\vert\varphi^\prime\rangle&=&
\sum^\infty_{n=-\infty}\int^{\lambda-\varepsilon}_{-\lambda+\varepsilon}
{dp\over2\pi}\ e^{i\left( p+J\right)\left(\delphi+2n\pi\right)}p\ ,\nonumber\\
\langle\varphi\vert J_\pm\vert\varphi^\prime\rangle&=&
\sum^\infty_{n=-\infty}\int^{\lambda-\varepsilon}_{-\lambda+\varepsilon}
{dp\over2\pi}\ e^{i\left( p+J\right)\left(\delphi+2n\pi\right)}
e^{\pm i\left(\overline\varphi+n\pi\right)}\sqrt{\lambda^2-p^2}\ ,
\label{gyouretsu}
\end{eqnarray}
where $\overline\varphi\equiv(\varphi+\varphi^\prime)/2$ and we have adopted
universal $\varepsilon$($0<\varepsilon<1/2$) to be able to have the same
integration range in all the matrix elements.
Thus the matrix element of the Hamiltonian (\ref{hamiltonian}) reads
\begin{eqnarray}
\langle\varphi\vert H\!\left( t\right)\vert\varphi^\prime\rangle
&=&\sum^\infty_{n=-\infty}
\int^{\lambda-\varepsilon}_{-\lambda+\varepsilon}{dp\over2\pi}\
e^{i\left( p+J\right)\left(\delphi+2n\pi\right)}\nonumber\\
&&\times\left[ h_3\!\left( t\right) p+\sqrt{\lambda^2-p^2}
\left\{ h_+\!\left( t\right)e^{-i\overline\varphi-in\pi}
+h_-\!\left( t\right) e^{i\overline\varphi+in\pi}\right\}\right]\ .
\label{hamyouso}
\end{eqnarray}
With these preliminaries the Feynman kernel,
\begin{equation}
K\left(\varphi_F,t_F;\varphi_I,t_I\right)\equiv
\lim_{N\to\infty}\langle\varphi_F\vert\left(1-i\delt H\!\left( N\right)\right)
\cdots\left(1-i\delt H\!\left( 1\right)\right)\vert\varphi_I\rangle\ ,
\end{equation}
where $H\!(j)\equiv H\!(t_j)$; $t_j\equiv t_I+j\delt$ with $\delt$
being given in (\ref{deltteigi}), can be expressed as
\begin{eqnarray}
K\left(\varphi_F,t_F;\varphi_I,t_I\right)
&=&\lim_{N\to\infty}\int^{2\pi}_0\prod^{N-1}_{i=1}d\varphi_i
\langle\varphi_N\vert\left(1\!-\!i\delt H\!\left( N\right)\right)
\vert\varphi_{N-1}\rangle\nonumber\\
&&\times\langle\varphi_{N-1}\vert\left(1\!-\!i\delt H\!\left( N\!-\!1\right)
\right)\vert\varphi_{N-2}\rangle\cdots\langle\varphi_1\vert
\left(1\!-\!i\delt H\!\left(1\right)\right)\vert\varphi_0\rangle
\Big\vert^{\varphi_N=\varphi_F}_{\varphi_0=\varphi_I}\nonumber\\
\!\!\!\!\!&=&\lim_{N\to\infty}\int^{2\pi}_0\prod^{N-1}_{i=1}d\varphi_i
\prod^N_{j=1}\langle\varphi_j\vert
\left(1-i\delt H\!\left( j\right)\right)\vert\varphi_{j-1}\rangle
\Big\vert^{\varphi_N=\varphi_F}_{\varphi_0=\varphi_I}\nonumber\\
\!\!\!\!&=&\lim_{N\to\infty}\int^{2\pi}_0\prod^{N-1}_{i=1}d\varphi_i
\int^{\lambda-\varepsilon}_{-\lambda+\varepsilon}\prod^N_{j=1}
\left({dp_j\over2\pi}\sum^\infty_{n_j=-\infty}\right)
\nonumber\\
&&\times\exp\Bigg[i\sum^N_{k=1}\Bigg\{\left(p_k+J\right)
\left(\delphi_k+2n_k\pi\right)
-\delt\Big\{h_3\!\left( k\right) p_k\label{kakunakaba}\\
&&+\left.\sqrt{\lambda^2\!-{p_k}^2}\left( h_+\!\left( k\right)
e^{-i\overline\varphi_k\!-\!in_k\pi}\!\!+\!h_-\!\left( k\right)
e^{i\overline\varphi_k\!+\!in_k\pi}\right)\right\}\Bigg]
\!\Bigg\vert^{\varphi_N=\varphi_F}_{\varphi_0=\varphi_I}\ ,\nonumber
\end{eqnarray}
with the help of the resolution of unity (\ref{phinoru}) and (\ref{hamyouso}),
where $\delphi_j$ are given in (\ref{delphiteigi}) and $\overline\varphi_j$ is
\begin{equation}
\overline\varphi_j\equiv{\varphi_j+\varphi_{j-1}\over2}\ ,
\end{equation}
and $O(\delt^2)$ term has been dropped to the final line.
Here a change of variables,
\begin{equation}
\cases{\displaystyle n^\prime_j=\sum^j_{k=1}n_k\ ;\
 \left(1\le j\le N\right) ,\cr
\displaystyle\varphi^\prime_j=\varphi_j+2n^\prime_j\pi
\ ;\ \left(1\le j\le N-1\right)\ ,\cr}\label{nphi}
\end{equation}
leads us to
\begin{eqnarray}
(\ref{kakunakaba})&=&
\lim_{N\to\infty}\sum^\infty_{n_N=-\infty}
\prod^{N-1}_{i=1}\sum^\infty_{n_i=-\infty}
\int^{2\left( n_i+1\right)\pi}_{2n_i\pi}d\varphi_i
\int^{\lambda-\varepsilon}_{-\lambda+\varepsilon}\prod^N_{j=1}
{dp_j\over2\pi}\nonumber\\
&&\times\exp\!\Bigg[ i\sum^N_{k=1}\!\Bigg\{\!\left( p_k+J\right)\delphi_k
-\delt\!\Big\{\! h_3\!\left( k\right) p_k\\
&&+\sqrt{\lambda^2\!-{p_k}^2}
\left( h_+\!\left( k\right) e^{-i\overline\varphi_k}\!+\!h_-\!\left( k\right)
e^{i\overline\varphi_k}\right)\Big\}\Bigg\}\Bigg]
\Bigg\vert^{\varphi_N=\varphi_F+2n_N\pi}_{\varphi_0=\varphi_I}\ ,\nonumber
\end{eqnarray}
where the primes have been dropped from $\varphi_j$ and $n_j$ and the relation,
\begin{equation}
\overline\varphi_j+n_j\pi=\overline\varphi^\prime_j-2n^\prime_{j-1}\pi\ ;\
e^{\pm i\left(\overline\varphi_j+n_j\pi\right)}
=e^{\pm i\overline\varphi^\prime_j}\ ,
\label{barn}
\end{equation}
has been utilized.
Writing
\begin{equation}
\sum^\infty_{n_j=-\infty}\int^{2\left( n_j+1\right)\pi}_{2n_j\pi}d\varphi=
\int^\infty_{-\infty}d\varphi\ ,
\end{equation}
we finally obtain
\begin{eqnarray}
&&K\left(\varphi_F,t_F;\varphi_I,t_I\right)=\sum^\infty_{n=-\infty}
K^{\left( n\right)}\left(\varphi_F,t_F;\varphi_I,t_I\right)\ ,\nonumber\\
&&K^{\left( n\right)}\left(\varphi_F,t_F;\varphi_I,t_I\right)\equiv
\lim_{N\to\infty}\int^\infty_{-\infty}\prod^{N-1}_{i=1}d\varphi_i
\int^{\lambda-\varepsilon}_{-\lambda+\varepsilon}\prod^N_{j=1}{dp_j\over2\pi}
\exp\left[ i\sum^N_{k=1}\left( p_k+J\right)\delphi_k\right.\nonumber\\
&&\left.\left.-i\delt\sum^N_{k=1}\left\{ h_3\!\left( k\right)p_k
+\sqrt{\lambda^2-{p_k}^2}\left( h_+\!\left( k\right)
e^{-i\overline\varphi_k}+h_-\!\left( k\right)e^{i\overline\varphi_k}\right)
\right\}\right]
\right\vert^{\varphi_N=\varphi_F+2n\pi}_{\varphi_0=\varphi_I}\nonumber\\
&&=e^{iJ\left(\varphi_F-\varphi_I+2n\pi\right)}
\lim_{N\to\infty}\int^\infty_{-\infty}\prod^{N-1}_{i=1}
d\varphi_i\int^{\pi-\delta}_\delta\prod^N_{j=1}
{\lambda\sin\theta_jd\theta_j\over2\pi}\label{atopcs}\\
&&\times\exp\left[ i\lambda\sum^N_{k=1}
\Bigg\{\cos\theta_k\delphi_k\!-\!\delt\left( h_3\!\left( k\right)\cos\theta_k
+\sin\theta_k\left( h_+\!\left( k\right)e^{-i\overline\varphi_k}+h_-\!\left(
k\right)e^{i\overline\varphi_k}\right)\right)\Bigg\}\right]\ ,
\nonumber
\end{eqnarray}
where we have introduced new variables by $p_j=\lambda\cos\theta_j$
with $\delta$ corresponding to $\varepsilon$.
This is the Nielsen-Rohrlich formula.
In view of (\ref{atopcs}), they called the phase space as a
``punctured sphere'' since we have a ``winding number, $n$'', in terms of
the infinite sum and the integration domain of $\theta$ is
$(\delta,\pi-\delta)$.

Next consider the same system in terms of a spin coherent state\cite{KS},
defined by
\begin{equation}
\left\vert\xi\right)\equiv e^{\xi J_+}\vert 0\rangle\ ,\
\vert\xi\rangle\equiv{1\over\left(\xi\vert\xi\right)^{1/2}}
\left\vert\xi\right)\ ,\ \left(\xi\in{\bf C}\right)\ ,
\end{equation}
or explicitly
\begin{equation}
\vert\xi\rangle={1\over\left(1+\vert\xi\vert^2\right)^J}
\sum^{2J}_{m=0}\xi^m{2J\choose m}^{1/2}\vert m\rangle\ .
\end{equation}
The inner product is
\begin{equation}
\langle\xi\vert\xi^\prime\rangle={\left(1+\xi^*\xi^\prime\right)^{2J}\over
\left(1+\vert\xi\vert^2\right)^J\left(1+\vert\xi^\prime\vert^2\right)^J}\ ,
\label{xinonaiseki}
\end{equation}
and the resolution of unity
\begin{equation}
{2J+1\over\pi}\int{d\xi^*d\xi\over\left(1+\vert\xi\vert^2\right)^2}
\vert\xi\rangle\langle\xi\vert\equiv\int d\mu\left(\xi,\xi^*\right)
\vert\xi\rangle\langle\xi\vert={\bf 1}_J\ ,\label{xinoru}
\end{equation}
is fulfilled, where
$d\xi^*d\xi\equiv d{\rm Re}\left(\xi\right) d{\rm Im}\left(\xi\right)$.
The matrix elements of generators read
\begin{eqnarray}
\langle\xi\vert J_3\vert\xi^\prime\rangle&=&-J{1-\xi^*\xi^\prime
\over1+\xi^*\xi^\prime}\langle\xi\vert\xi^\prime\rangle\ ,\nonumber\\
\langle\xi\vert J_+\vert\xi^\prime\rangle&=&J{2\xi^*\over
1+\xi^*\xi^\prime}\langle\xi\vert\xi^\prime\rangle\ ,\nonumber\\
\langle\xi\vert J_-\vert\xi^\prime\rangle&=&J{2\xi^\prime
\over1+\xi^*\xi^\prime}\langle\xi\vert\xi^\prime\rangle\ ,
\end{eqnarray}
giving
\begin{eqnarray}
\langle\xi\vert H\!\left( t\right)\vert\xi^\prime\rangle&=&
\langle\xi\vert h_3\!\left( t\right) J_3+h_-\!\left( t\right) J_++
h_+\!\left( t\right)J_-\vert\xi^\prime\rangle\nonumber\\
&=&J{-h_3\!\left( t\right)\left(1-\xi^*\xi^\prime\right)+2h_-\!\left( t\right)
\xi^*+2h_+\!\left( t\right)\xi^\prime\over
1+\xi^*\xi^\prime}\langle\xi\vert\xi^\prime\rangle\ .
\end{eqnarray}
The kernel in this case,
\begin{equation}
K\!\left(\xi_F,t_F;\xi_I,t_I\right)\equiv\lim_{N\to\infty}
\langle\xi_F\vert\left(1-i\delt H\!\left( N\right)\right)\cdots
\left(1-i\delt H\!\left(1\right)\right)\vert\xi_I\rangle\ ,
\end{equation}
turns out to be
\begin{eqnarray}
K\left(\xi_F,t_F;\xi_I,t_I\right)&=&\lim_{N\to\infty}\int\prod^{N-1}_{i=1}d\mu
\left(\xi_i,\xi_i^*\right)
\langle\xi_N\vert\left(1-i\delt H\!\left( N\right)\right)\vert\xi_{N-1}\rangle
\nonumber\\
&&\times\langle\xi_{N-1}\vert\left(1-i\delt H\!\left( N\!-\!1\right)\right)
\vert\xi_{N-2}\rangle\cdots
\langle\xi_1\vert\left(1-i\delt H\!\left(1\right)\right)\vert\xi_0\rangle
\Big\vert^{\xi_N=\xi_F}_{\xi_0=\xi_I}\nonumber\\
&=&\lim_{N\to\infty}\int\prod^{N-1}_{i=1}d\mu\left(\xi_i,\xi_i^*\right)
\exp\Bigg[ J\sum^N_{k=1}\Bigg\{
2\log{1+\xi_k^*\xi_{k-1}\over1+\vert\xi_k\vert^2}\nonumber\\
&&-i\delt{h_3\!\left( k\right)\left(1-\xi_k^*\xi_{k-1}\right)
+2h_-\!\left( k\right)
\xi_k^*+2h_+\!\left(
k\right)\xi_{k-1}\over1+\xi_k^*\xi_{k-1}}\Bigg\}\nonumber\\
&&\left.+\log\left(1+\vert\xi_N\vert^2\right)
-\log\left(1+\vert\xi_0\vert^2\right)\Bigg]
\right\vert^{\xi_N=\xi_F}_{\xi_0=\xi_I}\ .\label{sckernel}
\end{eqnarray}
In this case the phase space is $CP^1$, that is, {\em there is no hole
at all}.
(\ref{sckernel}) looks quite different from (\ref{atopcs}).

\section{Inspection of the ``Winding Number''}

The main difference of two formulae (\ref{atopcs}) and (\ref{sckernel})
is with or without the ``winding number'', so that in this section we discuss
quantum mechanics on $S^1$, where the nomenclature has first emerged,
and compare that to $SU(2)$.

Introduce canonical variables obeying
\begin{equation}
\left\{ p,\phi\right\}=-1\ ;
\ \left(-\infty<p,\phi<\infty\right)\ .
\end{equation}
In order to describe the mechanics on $S^1$,
$\phi$ is not a suitable coordinate since
on $S^1$ the periodicity with respect to $\phi$
must be respected.
Therefore consider $W=e^{i\phi}$ and
\begin{equation}
\left\{ W,p\right\}=iW\ .
\end{equation}
Quantum mechanically\cite{OK} thus
\begin{equation}
\left[\hat p,\hat W\right]=\hat W\ ,
\label{koukan}
\end{equation}
where $\hat p$ is self-adjoint, $\hat p^\dagger\!=\!\hat p$, and $\hat W$ is
unitary, $\hat W^\dagger\hat W\!=\!1$.
The representation is found by assuming
\begin{equation}
\hat p\vert\alpha\rangle=\alpha\vert\alpha\rangle\ ,\
\langle\alpha\vert\alpha\rangle=1\ ;\ \left(0\le\alpha<1\right)\ ,
\label{joutaia}
\end{equation}
so that
\begin{equation}
\hat p\vert m;\alpha\rangle=\left( m+\alpha\right)\vert m;\alpha\rangle\ ,\
\vert m;\alpha\rangle\equiv{\hat W}^m\vert\alpha\rangle\ ;\
\left( m\in{\bf Z}\right)\ ,
\end{equation}
which can be recognized by the commutation relation and (\ref{joutaia}).
Orthonormality
\begin{equation}
\langle m;\alpha\vert n;\alpha\rangle=\delta_{m,n}\ ,
\end{equation}
then the completeness,
\begin{equation}
\sum^\infty_{m=-\infty}\vert m;\alpha\rangle\langle m;\alpha\vert\equiv
{\bf 1}_\alpha\ ,\label{kanzensei}
\end{equation}
holds.
Each $\alpha$ thus specifies different Hilbert space, that is,
different ways of quantization on $S^1$.
While a $W$-diagonal representation,
\begin{equation}
\hat W\vert\phi\rangle=e^{i\phi}\vert\phi\rangle\ ;\
\left(0\le\phi<2\pi\right)\ ,
\end{equation}
can be constructed in terms of $\vert m;\alpha\rangle$ as
\begin{equation}
\vert\phi\rangle={1\over\sqrt{2\pi}}\sum^\infty_{m=-\infty}e^{im\phi}
\vert m;\alpha\rangle\left(\equiv\lim_{\varepsilon\to0}{1\over\sqrt{2\pi}}
\sum^\infty_{m=-\infty}e^{im\phi-\vert m\vert\varepsilon}
\vert m;\alpha\rangle\right)\ ,\label{joutaiphi}
\end{equation}
where regularization parameter $\varepsilon$ is needed for the convergence of
the infinite sum. (In the following however we suppress it.)
In view of (\ref{joutaiphi}) $\vert\phi\rangle$ is also regarded
as a periodic coherent state similar to (\ref{pcs}).
The inner product,
\begin{equation}
\langle\phi\vert\phi^\prime\rangle={1\over2\pi}
\sum^\infty_{m=-\infty}e^{im\left(\phi^\prime-\phi\right)}
=\sum^\infty_{n=-\infty}\delta\!\left(\phi^\prime-\phi+2\pi n\right)\ ,
\end{equation}
and the resolution of unity (completeness in this case),
\begin{equation}
\int^{2\pi}_0d\phi\vert\phi\rangle\langle\phi\vert={\bf 1}_\alpha\ ,
\label{kanzenseiphi}
\end{equation}
are easily convinced.

The Feynman kernel, when $\hat H=H\!(\hat p)$,
\begin{equation}
K\left(\phi_F,t_F;\phi_I,t_I\right)=\langle\phi_F\vert
e^{-i\left( t_F-t_I\right)\hat H}
\vert\phi_I\rangle=\lim_{N\to\infty}\langle\phi_F\vert\left(
1-i\delt H\!\left(\hat p\right)\right)^N\vert\phi_I\rangle\ ,
\end{equation}
becomes, by use of (\ref{kanzensei}) and (\ref{kanzenseiphi}), with
noting $\langle\phi\vert m;\alpha\rangle=e^{-im\phi}/\sqrt{2\pi}$, as
\begin{eqnarray}
K\left(\phi_F,t_F;\phi_I,t_I\right)&=&\lim_{N\to\infty}
\sum^\infty_{m_N=-\infty}\int^{2\pi}_0\prod^{N-1}_{i=1}
\left({d\phi_i\over2\pi}\sum^\infty_{m_i=-\infty}\right)\nonumber\\
&&\times\exp\left[ i\left\{\sum^N_{k=1}m_k\Delta\phi_k
-\delt H\left( m_k+\alpha\right)\right\}\right]
\Bigg\vert^{\phi_N=\phi_F}_{\phi_0=\phi_I}\nonumber\\
&=&\lim_{N\to\infty}\sum^\infty_{m_N=-\infty}\int^{2\pi}_0
\prod^{N-1}_{i=1}\left(d\phi_i\sum^\infty_{n_i=-\infty}\right)
\int^\infty_{-\infty}\prod^N_{j=1}{dp_j\over2\pi}\label{okk}\\
&&\times\exp\left[ i\sum^N_{k=1}\left\{\left(p_k-\alpha\right)
\left(\Delta\phi_k+2n_k\pi\right)-\delt H\!\left( p_k\right)\right\}\right]
\Bigg\vert^{\phi_N=\phi_F}_{\phi_0=\phi_I}\ ,\nonumber
\end{eqnarray}
where use has been made of the formula (\ref{kankeishiki}) by putting
$m_1=\infty$, $m_0=-\infty$ in the first and
of the shift $p_k\to p_k-\alpha$ in the second expression.
Now compare (\ref{okk}) with (\ref{kakunakaba}): apart from the integration
domain of $p_j$'s (as well as the form of the Hamiltonian)
the dependence of $\phi_j$ and $n_j$ is identical.
Therefore following the same procedure from (\ref{nphi}) to (\ref{barn}),
we obtain
\begin{eqnarray}
K\left(\phi_F,t_F;\phi_I,t_I\right)&=&\sum^\infty_{n=-\infty}
K^{\left( n\right)}\left(\phi_F,t_F;\phi_I,t_I\right)\nonumber\\
K^{\left( n\right)}\left(\phi_F,t_F;\phi_I,t_I\right)&\equiv&
\lim_{N\to\infty}e^{-i\alpha\left(\phi_F-\phi_I+2n\pi\right)}
\int^\infty_{-\infty}\prod^{N-1}_{i=1}d\phi_i
\int^\infty_{-\infty}\prod^N_{j=1}{dp_j\over2\pi}\nonumber\\
&&\times\exp\left[ i\sum^N_{k=1}\Bigg\{ p_k\Delta\phi_k-\delt H\!\left(
p_k\right)\Bigg\}\right]
\Bigg\vert^{\phi_N=\phi_F+2n\pi}_{\phi_0=\phi_I}\ .\label{ksone}
\end{eqnarray}
When $\alpha=0$, $K^{\left( n\right)}$ is the usual Feynman kernel in a
flat space except the boundary condition, $\phi_F=\phi_N+2n\pi$.
Significance of that is easily recognized; since on $S^1$ there is no
distinction in $n$ times around.
The quantum amplitude can therefore be obtained by summing up with
respect to $n$.
$n$ is designated as the {\em winding number}.

In view of formulae, in $S^1$ (\ref{ksone}) and $SU(2)$ cases (\ref{atopcs}),
there seems no difference for the appearance of the winding number,
but we should note the integration domain of $p_j$'s:
in $S^1$ case it stretches from $-\infty$ to $+\infty$
but in $SU(2)$ case the range is bounded.
According to the formula (\ref{kankeishiki}) the boundedness of
the $p$-integral does imply a finite rather than an infinite sum.
In (\ref{atopcs}) summation actually runs from $0$ to $2J$.
Therefore the ``winding number'' appeared in the $SU(2)$ periodic
coherent state expression is {\em superficial} so that {\em the phase space
cannot be regarded as having a nontrivial topology.}

As a final comment in this section, $SU(2)$ algebra can be realized
by means of (\ref{koukan}) such that
\begin{equation}
\hat J_3=\hat p\ ,\ \hat J_+=\hat W\sqrt{\lambda^2-\left(\hat p+1/2\right)^2}
\ ,\ \hat J_-=\sqrt{\lambda^2-\left(\hat p+1/2\right)^2}{\hat W}^\dagger
\ .\label{oksuni}
\end{equation}
With this while utilizing the previous recipe we can also construct
the NR formula\cite{KJ}.

\section{Relationship between Two Formulae}

We have recognized that the ``winding number'' in the NR formula (\ref{atopcs})
is superficial and that topology is not nontrivial.
The situation is thus similar to the formula
under the spin coherent state (\ref{sckernel}).
Therefore it might be possible to bridge the both expressions.
{}From now on we assume that the external field is constant for brevity's sake.

The trace of the kernel in the spin coherent state (\ref{sckernel}) reads as
\begin{eqnarray}
&&Z\equiv\int d\mu\left(\xi,\xi^*\right) K\left(\xi,T;\xi,0\right)
=\tr e^{-ihJ_3T}\nonumber\\
&&=\left.\lim_{N\to\infty}\int\prod^N_{i=1}d\mu\left(\xi_i,\xi_i^*\right)
\exp\left[ J\sum^N_{k=1}\left\{2\log{1+\xi_k^*\xi_{k-1}\over
1+\vert\xi_k\vert^2}+i\delt h{1-\xi_k^*\xi_{k-1}\over1+\xi_k^*\xi_{k-1}}
\right\}\right]\right\vert_{\xi_N=\xi_0}\nonumber\\
&&=\left. e^{ihJT}\lim_{N\to\infty}\int\prod^N_{i=1}
{\left(2J+1\right) d\xi_i^*d\xi_i\over\pi\left(1+\vert\xi_i\vert^2\right)^2}
\prod^N_{k=1}\left({1+e^{-i\delt h}{\xi_k}^*\xi_{k-1}\over
1+\vert\xi_k\vert^2}\right)^{2J}\right\vert_{\xi_N=\xi_0}\ ,
\label{atogcs}
\end{eqnarray}
where $d\mu(\xi,\xi^*)$ has been defined by (\ref{xinoru}) and $H=hJ_3$;
since Hamiltonian within the trace can always be
diagonalized by use of $SU(2)$ transformation.
By making a change of variables
\begin{equation}
\xi_k\mapsto\xi_ke^{-ik\delt h}\ ,
\end{equation}
(\ref{atogcs}) is reduced to a single integral such that
\begin{equation}
Z=e^{ihJT}\int d\mu\left(\xi,\xi^*\right)\langle\xi\vert\xi e^{-ihT}\rangle\ ,
\label{ichihen}
\end{equation}
yielding
\begin{equation}
Z={\sin\left(\left( J+1/2\right) hT\right)\over\sin\left( hT/2\right)}\ ,
\label{genmitsukekka}
\end{equation}
which is nothing but the $SU(2)$ character formula.

While the trace of the kernel under the periodic coherent state (\ref{atopcs})
is written as
\begin{eqnarray}
Z&=&\sum^\infty_{n=-\infty}Z^{\left( n\right)}\ ,\nonumber\\
Z^{\left( n\right)}&=&e^{i2n\pi J}\lim_{N\to\infty}
\int^{2\left( n+1\right)\pi}_{2n\pi}{d\varphi_N\over2\pi}
\int^\infty_{-\infty}\prod^{N-1}_{i=1}{d\varphi_i\over2\pi}
\int^\pi_0\prod^N_{j=1}\lambda\sin\theta_jd\theta_j\nonumber\\
&&\times\exp\left\{ i\lambda\sum^N_{k=1}\cos\theta_k
\left(\delphi_k-h\delt\right)\right\}\Bigg\vert_{\varphi_0=\varphi_N-2n\pi}\ ,
\label{atodiag}
\end{eqnarray}
where $\delta$, that is, $\varepsilon$ has been put zero as can be
recognized from the derivation (\ref{phiphidash}) and (\ref{gyouretsu}).
However the expression (\ref{atodiag}) can be cast into
another form as follows: first recall that the starting point was given by
\begin{equation}
Z=\tr e^{-ihJ_3T}=\lim_{N\to\infty}\int^{2\pi}_0\prod^N_{i=1}d\varphi_i
\prod^N_{j=1}\langle\varphi_j\vert\left(1-ih\delt J_3\right)\vert
\varphi_{j-1}\rangle\Big\vert_{\varphi_0=\varphi_N}\ ,\label{wkbmoto}
\end{equation}
where $\vert\varphi_j\rangle$ is the periodic coherent state (\ref{pcs}).
Instead of the resolution of unity (\ref{phinoru}), we introduce
\begin{equation}
{\bf 1}_J=\int^\infty_0{\left(2J+1\right) du\over\left(1+u\right)^2}
\sum^{2J}_{m=0}{2J\choose m}{u^m\over\left(1+u\right)^{2J}}
\vert m\rangle\langle m\vert\ ,\label{unoru}
\end{equation}
which can easily be convinced through the $u$-integral
(giving the Beta function $B(m+1,2J-m+1)$).
Use (\ref{unoru}) to find
\begin{eqnarray}
&&\langle\varphi_j\vert\left(1-ih\delt J_3\right)\vert\varphi_{j-1}\rangle
\nonumber\\
&=&\int^\infty_0{\left(2J+1\right)du_j\over2\pi\left(1+u_j\right)^2}
\sum^{2J}_{m=0}{2J\choose m}{u_j^m\over\left(1+u_j\right)^{2J}}
\langle\varphi_j\vert\left(1-ih\delt J_3\right)\vert m\rangle
\langle m\vert\varphi_{j-1}\rangle\ .\label{useki}
\end{eqnarray}
Performing the $u_j$-integrals after the substitution of (\ref{useki})
into (\ref{wkbmoto}) (and utilizing the formula (\ref{kankeishiki}))
we of course arrived at (\ref{atodiag}) again.
However we keep the $u$-integral intact and note the binomial formula
to obtain
\begin{equation}
\left(\ref{useki}\right)=\int^\infty_0{\left(2J+1\right)
du_j\over2\pi\left(1+u_j\right)^2}
e^{ihJ\delt}\left({1+u_je^{i\delphi_j-ih\delt}\over1+u_j}\right)^{2J}\ ,
\end{equation}
where we have used the second relation of (\ref{hyouji}) and
$\langle\varphi_j\vert m\rangle=e^{im\varphi_j}/\sqrt{2\pi}$.
Therefore
\begin{eqnarray}
&&Z=e^{ihJT}\lim_{N\to\infty}\int^{2\pi}_0\prod^N_{i=1}
{d\varphi_i\over2\pi}\int^\infty_0\prod^N_{j=1}
{\left(2J+1\right) du_j\over\left(1+u_j\right)^2}
\prod^N_{k=1}\left({1+u_ke^{-ih\delt+i\delphi_k}\over1+u_k}\right)^{2J}
\Bigg\vert_{\varphi_0=\varphi_N}\nonumber\\
&&=e^{ihJT}\lim_{N\to\infty}\int\prod^N_{i=1}
{\left(2J+1\right) d{\xi_i}^*d\xi_i\over\pi\left(1+\vert\xi_i\vert^2\right)^2}
\prod^N_{j=1}\left({1+{\xi_j}^*\xi_{j-1}e^{-ih\delt}\vert\xi_j/\xi_{j-1}\vert
\over1+\vert\xi_j\vert^2}\right)^{2J}\Bigg\vert_{\xi_0=\xi_N}
\ ,\label{taisho}
\end{eqnarray}
where use has been made of a change of variables
\begin{equation}
\xi_j=\sqrt{u_j}e^{-i\varphi_j}\ ,
\end{equation}
so that the boundary condition $\varphi_0=\varphi_N$ has been replaced to
$\xi_0=\xi_N$.
Now the final task is to show that (\ref{taisho}) is equivalent to
(\ref{atogcs}).
To this end introduce
\begin{equation}
K_x\!\left(\xi,\xi^\prime;\delt\right)\equiv
{\left(1+\xi^*\xi^\prime e^{-ih\delt}\vert\xi/\xi^\prime\vert^x\right)^{2J}
\over\left(1+\vert\xi\vert^2\right)^J\left(1+\vert\xi^\prime\vert^2\right)^J}\
,
\end{equation}
which satisfies the kernel property:
\begin{equation}
\int d\mu\left(\xi_j,{\xi_j}^*\right) K_x\!\left(\xi_{j+1},\xi_j;\delt\right)
K_x\!\left(\xi_j,\xi_{j-1};\delt\right)
=K_x\!\left(\xi_{j+1},\xi_{j-1};2\delt\right)\ .\label{tprop}
\end{equation}
Note that
\begin{equation}
\left(\ref{taisho}\right)=e^{ihJT}\lim_{N\to\infty}
\int\prod^N_{i=1}d\mu\left(\xi_i,\xi_i^*\right)\prod^N_{j=1}
K_{x=1}\!\left(\xi_j,\xi_{j-1};\delt\right)\Bigg\vert_{\xi_0=\xi_N}\ .
\label{then}
\end{equation}
We now show that the quantity,
\begin{equation}
Z_x\equiv e^{ihJT}\lim_{N\to\infty}\int\prod^N_{i=1}
d\mu\left(\xi_i,\xi_i^*\right)\prod^N_{j=1}
K_x\!\left(\xi_j,\xi_{j-1};\delt\right)\Bigg\vert_{\xi_0=\xi_N}\ .
\end{equation}
is independent of $x$ and furthermore equal to $Z$ (\ref{ichihen});
since by use of (\ref{tprop}),
\begin{eqnarray}
Z_x&=&\left.e^{ihJT}\lim_{N\to\infty}\int d\mu\left(\xi_N,{\xi_N}^*\right)
{\left(1+{\xi_N}^*\xi_0e^{-ihN\delt}\left\vert\xi_N/\xi_0\right\vert^x
\right)^{2J}\over\left(1+\vert\xi_N\vert^2\right)^J\left(1+\vert\xi_0\vert^2
\right)^J}\right\vert_{\xi_0=\xi_N}\nonumber\\
&=&e^{ihJT}\int d\mu\left(\xi,\xi^*\right){\left(1+\xi^*\xi
e^{-ihT}\right)^{2J}
\over\left(1+\vert\xi\vert^2\right)^{2J}}\nonumber\\
&=&e^{ihJT}\int d\mu\left(\xi,\xi^*\right)\langle\xi\vert\xi e^{iT}\rangle
=Z=\left(\ref{ichihen}\right)\ ,
\end{eqnarray}
where use has been made of (\ref{xinonaiseki}) in the final expression.
Therefore (\ref{taisho}), that is, (\ref{then}) is equivalent to
(\ref{ichihen}).
We thus confirm a connection of two formulae:
the NR and the spin coherent state.

\section{The WKB Approximation}

According to the observation of the foregoing section there exists
equivalence between the two formulae of NR (\ref{atodiag}) and of the
spin coherent state (\ref{atogcs}).
In the latter case, it has been shown that the WKB approximation
gives the exact result\cite{FKSF}.
While in the former case it seems impossible for the WKB approximation
in terms of $p$ and $\varphi$ as was mentioned in the introduction.
Therefore we start with (\ref{atodiag}), that is, in terms of
$\theta$ and $\varphi$.

The action is read as
\begin{equation}
S=\lambda\sum^N_{k=1}\cos\theta_k\left(\delphi_k-h\delt\right)\ ,\label{sayou}
\end{equation}
and the WKB approximation is realized when $\lambda\to\infty$.
The stationary phase condition is given by
\begin{equation}
\cases{\sin\theta_j\left(\delphi_j-h\delt\right)=0\ ,\cr
\cos\theta_{j+1}-\cos\theta_j=0\ ;\
\left(\theta_{N+1}\equiv\theta_1\right)\ .}
\end{equation}
As was mentioned in (\ref{eqm}), $\delphi_j-h\delt=0$ cannot be accepted
because of the boundary condition, $\varphi_0=\varphi_N-2n\pi$.
Instead solutions,
\begin{equation}
\cases{\theta_j=0\ ,\cr\theta_j=\pi\ ,\cr}
\end{equation}
with $\varphi$'s being arbitrary, can be adopted.

First we consider the $\theta=0$ case:
by putting $\theta_j\to x_j/\sqrt{\lambda}$,
$Z^{\left( n\right)}$ (\ref{atodiag}) becomes
\begin{eqnarray}
Z^{\left( n\right)}\!\left(\theta=0\right)&\equiv&e^{i2n\pi J}\lim_{N\to\infty}
\int^{2\left( n+1\right)\pi}_{2n\pi}{d\varphi_N\over2\pi}
\int^\infty_{-\infty}\prod^{N-1}_{i=1}{d\varphi_i\over2\pi}\nonumber\\
&&\times\prod^N_{j=1}\Bigg[ e^{i\lambda\left(\delphi_j-h\delt\right)}
\int^\infty_0dx_j\ x_j\Bigg\{\sum^\infty_{m_j=0}{\left(-1\right)^{m_j}\over
\left(2m_j+1\right)!}{x_j^{2m_j}\over\lambda^{m_j}}\Bigg\}
e^{-i\left(\delphi_j-h\delt-i\varepsilon\right) x_j^2/2}\nonumber\\
&&\times\exp\Bigg\{ i\lambda\left(\delphi_j-h\delt\right)\sum^\infty_{m_j=2}
{\left(-1\right)^{m_j}\over\left(2m_j+1\right)!}{x_j^{2m_j}\over\lambda^{m_j}}
\Bigg\}\Bigg]\Bigg\vert_{\varphi_0=\varphi_N-2n\pi}\ ,\label{wkbpcs}
\end{eqnarray}
where expansions with respect to $x_j/\sqrt{\lambda}$ have been made
and $\varepsilon$ that will be put zero finally has been introduced
in order to assure the convergence of the Gaussian integral.
However in view of the integral (\ref{ippan}), $Z$ (\ref{atodiag}) does
belong to the class; by regarding $dx$ and $f\!(x)$ as $d\theta_j$ and
$\cos\theta_j$ respectively.
That is, in (\ref{atodiag}) the WKB approximation gives the exact result.
To see this consider the leading term of (\ref{wkbpcs}):
\begin{eqnarray}
Z^{\left( n\right)}_0\!\left(\theta=0\right)&\equiv&e^{i2n\pi J}
\int^{2\left( n+1\right)\pi}_{2n\pi}{d\varphi_N\over2\pi}
\int^\infty_{-\infty}\prod^{N-1}_{i=1}{d\varphi_i\over2\pi}\nonumber\\
&&\times\prod^N_{j=1}\Bigg\{ e^{i\lambda\left(\delphi_j-h\delt\right)}
\int^\infty_0dx_j\ x_je^{-{i\over2}\left(\delphi_j-h\delt-i\varepsilon
\right) x_j^2}\Bigg\}\Bigg\vert_{\varphi_0=\varphi_N-2n\pi}\nonumber\\
&=&e^{i2n\pi\left( J+\lambda\right)-i\lambda hT}
\lim_{N\to\infty}\int^{2\left( n+1\right)\pi}_{2n\pi}{d\varphi_N\over2\pi}
\int^\infty_{-\infty}\prod^{N-1}_{i=1}{d\varphi_i\over2\pi}\nonumber\\
&&\times\prod^N_{j=1}{1\over i\left(\delphi_j-h\delt-i\varepsilon\right)}
\Bigg\vert_{\varphi_0=\varphi_N-2n\pi}\ .
\end{eqnarray}
Taking account of
\begin{equation}
\int^\infty_{-\infty}{d\varphi_j\over2\pi i}{1\over\delphi_{j+1}-\delt
-i\varepsilon}{1\over\delphi_j-\delt-i\varepsilon}
={1\over\varphi_{j+1}-\varphi_{j-1}-2\delt-i\varepsilon}\ ,
\end{equation}
we can perform the $\varphi$-integral to find
\begin{eqnarray}
Z^{\left( n\right)}_0\!\left(\theta=0\right)&=&
e^{i2n\pi\left( J+\lambda\right)-i\lambda hT}
\int^{2\left( n+1\right)\pi}_{2n\pi}{d\varphi_N\over2\pi i}
{1\over2n\pi-hT}\nonumber\\
&=&{e^{i2n\pi\left( J+\lambda\right)-i\lambda hT}\over
i\left(2n\pi-hT\right)}\ .
\end{eqnarray}
In a similar manner, the contribution from $\theta_j=\pi$ reads
\begin{equation}
Z^{\left( n\right)}_0\!\left(\theta=\pi\right)=
-{e^{i2n\pi\left( J-\lambda\right)+i\lambda hT}\over i\left( 2n\pi-hT\right)}\
{}.
\end{equation}
Thus the total contribution is
\begin{eqnarray}
Z_{\rm WKB}&=&
\sum^\infty_{n=-\infty}\left( Z^{\left( n\right)}_0\!\left(\theta=0\right)+
Z^{\left( n\right)}_0\!\left(\theta=\pi\right)\right)\nonumber\\
&=&\sum^\infty_{n=-\infty}\left\{{e^{i2n\pi\left( J+\lambda\right)
-i\lambda hT}\over i\left(2n\pi-hT\right)}-{e^{i2n\pi\left( J-\lambda\right)
+i\lambda hT}\over i\left(2n\pi-hT\right)}\right\}\nonumber\\
&=&{\sin\left(\left( J+1/2\right) hT\right)\over\sin\left( hT/2\right)}\ ,
\label{wkbkekka}
\end{eqnarray}
where use has been made of the formula
\begin{equation}
\sum^\infty_{n=-\infty}{e^{i2n\pi\varepsilon}\over2n\pi+\varphi}=
{e^{i\left(1/2-\varepsilon\right)\varphi}\over2\sin{\varphi\over2}}\ ;\
\left(0<\varepsilon<1\right)\ .
\end{equation}
(Note that in view of (\ref{lteigi}), $J\pm\lambda=\pm1/2+{\rm mod}\ {\bf Z}$.)
Of course this result (\ref{wkbkekka}) coincides with the exact result
(\ref{genmitsukekka}).

\section{Discussion}

In this paper we have clarified the relationship between
the Nielsen-Rohrlich and the spin coherent state path integral
formulae for spin $SU(2)$.
In spite of a discrepancy of the appearance there is no
difference between them.
Therefore the first observation, by Nielsen and Rohrlich,
that the phase space of spin was given by a punctured sphere
is not correct.
Consequently the WKB approximation to the Nielsen-Rohrlich
formula has been shown to give an exact result as was done for
the formula under the spin coherent state\cite{FKSF}.

Our discussion was made upon the trace form of the kernel;
(\ref{atopcs}) and (\ref{sckernel}).
It would be better to be able to make a connection between themselves.
In other words, a direct relationship could be confirmed
if some change of variables, from $(\theta_j,\varphi_j)$
to $(\xi^*_j,\xi_j)$, would be found.
Classically
\begin{equation}
\xi=e^{i\varphi}\tan{\theta\over2}\ ,\label{kotenhenkan}
\end{equation}
is a desired relation;
since by putting $\delt\to0$ naively (and $h=0$) in (\ref{sckernel})
the exponent reads
\begin{equation}
-2J{\xi^*\dot\xi\over1+\xi^*\xi}\ ,
\end{equation}
which becomes by means of (\ref{kotenhenkan}) as
\begin{equation}
-2J{\xi^*\dot\xi\over1+\xi^*\xi}=
iJ\cos\theta\ \dot\varphi+{\rm total\ derivatives}\ ,
\end{equation}
which is nothing but the exponent of (\ref{atopcs}).
However quantum mechanically the task is tough
since relations should be dictated in terms of difference
instead of differential form.
Of course a connection could be found
if a case would happen to be
described by canonical transformation\cite{FK},
but apparently our case is not.

Although the topology of the NR formula is not the punctured sphere,
the nomenclature is still captivating in order to memorize
the formula (\ref{atopcs}):
with an infinite sum as well as the integration domain
being given $\delta\le\theta\le\pi-\delta$.
Both formulae are equivalent so that a suitable one should be
chosen in a given situation.

\vspace{2cm}
\begin{center}{\bf\large Acknowledgments}\end{center}

The authors would thank to K. Fujii for fruitful discussions.

\vspace{2cm}
\begin{center}{\Large\bf Appendix}\end{center}
\appendix
\section{Some WKB Exact Integral}
In this appendix we show that in some integral the stationary phase (WKB)
approximation gives the exact result: consider
\begin{equation}
I=\int^b_adxf^\prime\left( x\right) e^{-igf\left( x\right)}\ ,\label{ippan}
\end{equation}
with conditions,
\begin{equation}
f^\prime\left( a\right)=f^\prime\left( b\right)=0\ ,\
f^{\prime\prime}\left( a\right)\ne0\ ,\ f^{\prime\prime}\left( b\right)\ne0\ ,
\ g>0\ .
\end{equation}
The integral, of course, can trivially be performed to be
\begin{equation}
I={1\over ig}\left( e^{-igf\left( a\right)}-e^{-igf\left( b\right)}\right)\ ,
\label{jimei}
\end{equation}
whose result, however, can be obtained also from the WKB approximation when
$g\to\infty$:
let us take the case $x=a$.
Expand all quantities around $x=a$ by putting
\begin{equation}
x=a+{y\over\sqrt{g}}
\end{equation}
to find
\begin{eqnarray}
I\mapsto I\!\left( x=a\right)&\equiv&
e^{-igf\left( a\right)}\int^\infty_0{dy\over\sqrt{g}}\sum^\infty_{m=0}
{1\over\left( m+1\right)!}f^{\left( m+2\right)}\left( a\right)
\left({y\over\sqrt{g}}\right)^{m+1}
\exp\left(-{i\over2}f_\varepsilon^{\prime\prime}\left( a\right) y^2\right)
\nonumber\\
&&\times\exp\left\{-ig\sum^\infty_{n=0}
{1\over\left( n+3\right)!}f^{\left( n+3\right)}\left( a\right)
\left({y\over\sqrt{g}}\right)^{n+3}
\right\}\ ,\label{iateigi}
\end{eqnarray}
where $f_\varepsilon^{\prime\prime}(a)\equiv f^{\prime\prime}(a)-i\varepsilon$
with $\varepsilon$ assuring the convergence of the integral.
The upper limit of the integral has been set to infinity since that
was given by $\sqrt{g}(b-a)$.
(\ref{iateigi}) is further rewritten to
\begin{eqnarray}
I\!\left( x=a\right)&=&e^{-igf\!\left( a\right)}\sum^\infty_{m=0}
{f^{\left(m+2\right)}\!\left( a\right)\over\left( m+1\right)!}
\sum^\infty_{k=0}{\left(-i\right)^k\over k!}\prod^k_{i=1}
\left\{\sum^\infty_{n_i=1}{f^{\left( n_i+3\right)}\!\left( a\right)
\over\left( n_i+3\right)!}\right\}\nonumber\\
&&\times\left({1\over\sqrt{g}}\right)^{m+2+k+\sum^k_{i=1}n_i}
\int^\infty_0dy\ y^{m+1+3k+\sum^k_{i=1}n_i}
e^{-if_\varepsilon^{\prime\prime}\left( a\right) y^2/2}\ .
\end{eqnarray}
Utilize an identity
\begin{eqnarray}
1&=&\sum^\infty_{N=2}\delta_{N,m+2+k+\sum^k_{i=1}n_i}\nonumber\\
&=&\sum^\infty_{N=2}\oint{dz\over2\pi i}z^{m+2+k-N+\sum^k_{i=1}n_i}\ ,
\end{eqnarray}
(noting that $m+2+k+\sum^k_{i=1}n_i\ge2$) to find
\begin{eqnarray}
I\!\left( x=a\right)&=&e^{-igf\left( a\right)}\sum^\infty_{N=2}
\left({1\over\sqrt{g}}\right)^N\oint{dz\over2\pi i}
\left\{\sum^\infty_{m=0}{f^{\left( m+2\right)}\!\left( a\right)
\over\left( m+1\right)!}z^{m+1}\right\}
\sum^\infty_{k=0}{\left(-i\right)^k\over k!}\nonumber\\
&&\times\prod^k_{n_i=0}\left\{\sum^\infty_{n_i=0}{f^{\left( n_i+3\right)}
\!\left( a\right)
\over\left( n_i+3\right)!}z^{n_i+1}\right\}
\int^\infty_0dy\ y^{N+2k-1}
e^{-if_\varepsilon^{\prime\prime}\left( a\right) y^2/2}\nonumber\\
&=&e^{-igf\left( a\right)}\sum^\infty_{N=2}\left({1\over\sqrt{g}}\right)^N
\sum^\infty_{k=0}{1\over k!}\oint{dz\over2\pi i}z^{-N}
f^{\prime}\!\left( z+a\right)\left(-ih\!\left( z\right)\right)^k\nonumber\\
&&\times\int^\infty_0{dt\over2}t^{N/2+k-1}
e^{-if_\varepsilon^{\prime\prime}\left( a\right) t/2}\ ,
\end{eqnarray}
where
\begin{eqnarray}
h\left( z\right)&\equiv&\sum^\infty_{n=0}{1\over\left( n+3\right)!}
f^{\left( n+3\right)}\left( a\right) z^{n+1}\nonumber\\
&=&{1\over z^2}\left\{ f\left( z+a\right)-f\left( a\right)-{1\over2}
f^{\prime\prime}\left( a\right) z^2\right\}\ ,\label{hteigi}\\
f^\prime\!\left( z+a\right)&=&\sum^\infty_{m=0}
{f^{\left( m+2\right)}\!\left( a\right)\over\left( m+1\right)!}z^{m+1}\ ,
\end{eqnarray}
and we have made a change of variables, $y\mapsto t=y^2$, in the
final expression.
Taking account of
\begin{equation}
\int^\infty_0dt\ t^\alpha e^{-\beta t}={\Gamma\left(\alpha+1\right)\over
\beta^{\alpha+1}}\ ,\ \left({\rm Re}\beta>0\right)\ ,
\end{equation}
we perform the $t$-integral to find
\begin{eqnarray}
I\!\left( x=a\right)&=&e^{-igf\left( a\right)}\sum^\infty_{N=2}
\left({1\over\sqrt{g}}\right)^N
\sum^\infty_{k=0}{1\over k!}\left(-i\right)^k{1\over2}
\Gamma\left({N\over2}+k\right)\nonumber\\
&&\times\left({2\over if^{\prime\prime}\left( a\right)}\right)^{N/2+k}
\oint{dz\over2\pi i}z^{-N}\left\{ h\left( z\right)\right\}^k
\left\{ z^2h^\prime\left( z\right)+2zh\left( z\right)+zf^{\prime\prime}
\left( a\right)\right\}\nonumber\\
&=&{1\over2}e^{-igf\left( a\right)}\sum^\infty_{N=2}
\left({1\over\sqrt{g}}\right)^N\left( A_N+B_N\right)\ ,
\end{eqnarray}
where we have used (\ref{hteigi}) and
\begin{eqnarray}
A_N\!\!&\!\!\equiv\!\!&\!\!\sum^\infty_{k=0}{\left(-i\right)^k\over k!}
\Gamma\!\left({N\over2}
+k\right)\left({2\over if^{\prime\prime}\left( a\right)}\right)^{N/2+k}
\!\!\!\oint{dz\over2\pi i}\left\{ h\!\left( z\right)\right\}^k
\left\{{1\over z^{N-2}}h^\prime\!\left( z\right)
\!+\!{1\over z^{N-1}}2h\!\left( z\right)\right\}\ ,\nonumber\\
B_N&\equiv&\sum^\infty_{k=0}{\left(-i\right)^k\over k!}
\Gamma\!\left({N\over2}+k\right)
f^{\prime\prime}\!\left( a\right)
\left({2\over if^{\prime\prime}\left( a\right)}\right)^{N/2+k}
\oint{dz\over2\pi i}{1\over z^{N-1}}\left\{ h\left( z\right)\right\}^k
\ .
\end{eqnarray}
Noting that $h(z)$ and $h^\prime(z)$ are regular at $z=0$ and $h(0)=0$,
we obtain when $N=2$
\begin{equation}
A_2=0\ ,\ B_2={2\over i}\ .
\end{equation}
But when $N\ge3$,
\begin{eqnarray}
A_N&=&\sum^\infty_{k=0}\left(-i\right)^k
\left({2\over if^{\prime\prime}\left( a\right)}\right)^{N/2+k}
{\Gamma\!\left( N/2+k\right)\left( N+2k\right)
\over\left( k+1\right)!\left( N-2\right)!}
\left({d\over dz}\right)^{N-2}\!\!\left\{ h\!\left( z\right)\right\}^{k+1}
\Bigg\vert_{z=0}\nonumber\\
&=&-{2\over i}\sum^\infty_{k=1}\left(-i\right)^k
\left({2\over if^{\prime\prime}\left( a\right)}\right)^{N/2+k-1}
{\Gamma\!\left( N/2+k\right)\over k!\left( N-2\right)!}
\left({d\over dz}\right)^{N-2}\!\!\left\{ h\!\left( z\right)\right\}^k
\Bigg\vert_{z=0}\ ,
\end{eqnarray}
where use has been made of $(N+2k)\Gamma(N/2+k)=2\Gamma(N/2+k+1)$
and the shift $k\to k-1$ to the final expression.
And
\begin{eqnarray}
B_N&=&\sum^\infty_{k=0}\left(-i\right)^k\Gamma\!\left({N\over2}
+k\right) f^{\prime\prime}\!\left( a\right)
\left({2\over if^{\prime\prime}\left( a\right)}\right)^{N/2+k}
{1\over k!\left( N-2\right)!}
\left({d\over dz}\right)^{N-2}\!\!\left\{ h\!\left( z\right)\right\}^k
\Bigg\vert_{z=0}\nonumber\\
&=&{2\over i}\sum^\infty_{k=1}\left(-i\right)^k
\left({2\over if^{\prime\prime}\left( a\right)}\right)^{N/2+k-1}
{\Gamma\!\left( N/2+k\right)\over k!\left( N-2\right)!}
\left({d\over dz}\right)^{N-2}\!\!\left\{ h\!\left( z\right)\right\}^k
\Bigg\vert_{z=0}\ ,
\end{eqnarray}
where $k=0$ term does not have any contribution so that
the sum starts from $k=1$ in the final line.
Apparently
\begin{equation}
A_N+B_N=0\ .
\end{equation}
Thus there survives only the leading order yield
\begin{eqnarray}
I\!\left( x=a\right)&=&{1\over2}e^{-igf\left( a\right)}
\left({1\over\sqrt{g}}\right)^2{2\over i}\nonumber\\
&=&{1\over ig}e^{-igf\left( a\right)}\ .
\end{eqnarray}
In a similar manner, we find $I\!(x=b)=-e^{-igf\left( b\right)}/ig$ to obtain
\begin{equation}
I_{\rm WKB}\equiv I\!\left( x=a\right)+I\!\left( x=b\right)=
{1\over ig}\left( e^{-igf\left( a\right)}-e^{-igf\left( b\right)}\right)
=\left(\ref{jimei}\right)\ .
\end{equation}
Therefore the leading order of the stationary phase approximation
gives the exact result.

\end{document}